\documentclass[nofootinbib,reprint,
longbibliography
]{revtex4-1}
\usepackage{amsmath}
\usepackage{hyperref}
\hypersetup{
	colorlinks = true,
	citecolor = {blue},
	urlcolor = {blue}
}
\usepackage{bm}

\newcommand{\be}{\begin{equation}}
\newcommand{\ee}{\end{equation}}

\begin{document}


\title{The Quantum Focusing Conjecture Has Not Been Violated}
\author{Stefan Leichenauer}
\email{sleichen@berkeley.edu}
\affiliation{Department of Physics,
University of California, Berkeley, CA 94720, U.S.A.} 
\affiliation{Lawrence Berkeley National Laboratory, Berkeley, CA 94720, U.S.A.}

\begin{abstract}
Recent work of Fu, Koeller, and Marolf shows that in $d\geq 5$ dimensions a nonzero Gauss-Bonnet coupling of either sign can lead to a pointwise violation of the Quantum Focusing Conjecture. This violation is due to the classical geometric terms appearing in the QFC. Since those geometric terms are properly understood as expectation values of operators in an effective field theory, we argue that they are only well-defined when smeared over a region at least as large as the cutoff scale of the theory (which may be the Planck scale). We find that this smearing prescription removes the pointwise violation found by Fu et al.. We comment on the relationship to similar issues encountered in the study of Entanglement Wedge Nesting in holography.
\end{abstract}


\maketitle

\section{Introduction}\label{sec-intro}

The Quantum Focusing Conjecture (QFC) was proposed in \cite{Bousso:2015mna} as a quantum extension of the classical focusing theorem in General Relativity. Classically, focusing states that the expansion $\theta$ of a null congruence of geodesics is nonincreasing along the geodesic flow. This follows from the null curvature condition, which in turn follows from the null energy condition in GR. Classical focusing is a cornerstone of many important results in classical gravity~\cite{Penrose:1964wq, Hawking:1971tu, Hawking:1991nk, Friedman:1993ty}. The null curvature condition, and hence classical focusing, may be violated by quantum effects~\cite{Epstein:1965zza}. But there is a suitable replacement: quantum focusing, which says that a a certain function $\Theta$, called the quantum expansion and to be defined below, is nonincreasing along a geodesic flow.

The QFC elevates quantum focusing to a principle of semiclassical quantum gravity, more fundamental than any energy condition, from which other results should be derived. With the QFC as an assumption, many quantum analogues of the classical gravity results can be proven~\cite{Wall:2009wi, C:2013uza, Bousso:2015eda, Akers:2016ugt}. In addition, the nongravitational limit of the QFC results in the Quantum Null Energy Condition (QNEC)~\cite{Bousso:2015mna}, which has been independently proved in multiple contexts~\cite{Bousso:2015wca, Koeller:2015qmn, Akers:2016ugt}. Given this success, one should have confidence that the QFC is a true statement about semicalssical gravity.

The extremely interesting recent paper of Fu, Koeller, and Marolf shows that a naive formulation of the QFC is incorrect~\cite{Fu:2017lps}. They demonstrate convincingly that pointwise quantum focusing can be violated in theories with nonzero Gauss--Bonnet couplings in $d\geq 5$. In effective field theory such a coupling is generic, and so this issue must be addressed. Motivated by that example, we revisit the formulation of the QFC and find a resolution. One cannot treat the geometry purely classically: the geometric quantities must be appropriately averaged over regions at least as large as the cutoff scale in order to be well-defined, and this allows us to avoid a QFC violation. The key point is that we have to consistently treat the geometry within the rules of effective field theory. In Sec.~\ref{sec-QFC} we will review the formulation of the QFC, in Sec.~\ref{sec-problem} we review the violation pointed by Fu et al., and then in Sec.~\ref{sec-solution} we demonstrate that the appropriate operator smearing prevents the violation. We will conclude in Sec.~\ref{sec-disc} with a brief discussion of some related results, including a similar issue that arises in the study of Entanglement Wedge Nesting in holography.

\section{Statement of the QFC}\label{sec-QFC}

In this section we review the statement of the QFC. At the end we will emphasize the operator smearing aspects which lead to the resolution of the Fu et al. violation.

The formulation of the QFC begins with a codimension-two Cauchy-splitting surface $\Sigma$ and a null vector field $k^\mu$ normal to $\Sigma$, so that $\Sigma$ can be usefully thought of as a cut of the null surface $N$ generated by $k^\mu$. For future use, let $y$ be a set of coordinates on $\Sigma$. Associated to any codimension-two Cauchy-splitting surface such as $\Sigma$ we have the generalized entropy $S_{\rm gen}[\Sigma]$, defined by
\be
S_{\rm gen} = S_{\rm grav} + S_{\rm out}.
\ee
Here $S_{\rm grav}$ consists of local geometric terms integrated over $\Sigma$, while $S_{\rm out}$ is the renormalized von Neumann entropy of the region outside\footnote{We could have also used the von Neumann entropy of the inside of $\Sigma$. In a pure state the two are equal.} of $\Sigma$. The geometric terms in $S_{\rm grav}$ can in principle be obtained from the low-energy effective action for the metric, and are known explicitly for a wide class of actions. $S_{\rm gen}$ is well-defined because the renormalization scale dependence of $S_{\rm out}$ is canceled by that of the couplings appearing in $S_{\rm grav}$.

For a fixed semiclassical state, we can treat $S_{\rm gen}$ as a functional of the surface $\Sigma$. In particular, we will want to consider variations of $S_{\rm gen}$ as $\Sigma$ is deformed within the null surface $N$. Consider the one-parameter family $\Sigma(\lambda)$ of cuts of $N$, defined by flowing along the null congruence generated by $f(y)k^\mu$ for an affine parameter $\lambda$, beginning with $\Sigma$ at $\lambda=0$ and with an arbitrary function $f(y)\geq0$. Then we can define the quantum expansion $\Theta[\Sigma, y]$ through the equation
\be\label{Sgen'}
\left.\frac{dS_{\rm gen}}{d\lambda}\right|_{\lambda =0} = \frac{1}{4G}\int_\Sigma d^{d-2}y\sqrt{h}~\Theta[\Sigma,y]f(y).
\ee
Note that we have factored out the induced volume factor $\sqrt{h}$, and that $\Theta$ itself is independent of $f(y)$. Equivalently, using functional derivative notation we would write
\be
\Theta[\Sigma,y] = \frac{4G}{\sqrt{h}} \frac{\delta S_{\rm gen}}{\delta \Sigma(y)}.
\ee
The leading term in $S_{\rm grav}$ is $A[\Sigma]/4G$, which means that
\be
\Theta[\Sigma,y] = \theta(y) + \cdots,
\ee
where $\theta = \nabla_\mu k^\mu$ is the expansion of the null congruence generated by $k^\mu$. Thus we see explicitly how the quantum expansion $\Theta$ generalizes the classical expansion $\theta$.

The QFC postulates the inequality
\be\label{QFC}
0 \geq \frac{d\Theta[\Sigma,y]}{d\lambda} = \int d^{d-2}y' ~\frac{\delta\Theta[\Sigma,y]}{\delta \Sigma(y')}f(y').
\ee
Since this must be true for any nonnegative $f(y)$, it must be that $\delta\Theta[\Sigma,y]/\delta \Sigma(y')$ itself is nonpositive. Since $S_{\rm grav}$ is a local integral of geometric quantities on $\Sigma$, it will only contribute a $\delta$-function term to $\delta\Theta[\Sigma,y]/\delta \Sigma(y')$. $S_{\rm out}$ can also contribute to a $\delta$-function term, though it is much more difficult to analyze. The QFC requires that the total coefficient of the $\delta$-function be nonpositive. The contributions from $y\neq y'$ all originate with $S_{\rm out}$, and can be shown to be nonpositive using strong subadditivity~\cite{Bousso:2015mna}.

There are two facts that will come into play later. The first, and most important, is that one must remember that the contributions to $\delta\Theta[\Sigma,y]/\delta \Sigma(y')$ coming from $S_{\rm grav}$ are actually the expectation values of local operators in an effective quantum field theory. As such, to be well-defined they must be smeared over a distance scale at least as large as the cutoff scale of the theory. Since $\Theta$ is associated to the surface $\Sigma$, we will interpret these operators as surface operators and hence will only require that they be smeared along $\Sigma$. A more detailed understanding of the theory may reveal that more smearing is necessary to tame all divergences, but we will be conservative and only smear along $\Sigma$. That will turn out to be enough.

The second fact is that the function $f(y)$ appearing in \eqref{QFC}, while formally arbitrary, cannot reasonably be taken to have support smaller than the cutoff scale. This is because $S_{\rm gen}$ and its variations, which are effective field theory quantities, can only be reliably computed for surfaces which do not have sharp features on scales of order the cutoff. Since the function $f(y)$ cannot be arbitrarily well-localized, it means that the coefficient of the $\delta$-function in $\delta\Theta[\Sigma,y]/\delta \Sigma(y')$ cannot be perfectly isolated from the non-local terms.\footnote{Similar reasoning applied to \eqref{Sgen'} provides additional justification for smearing the local geometric terms along the surface $\Sigma$.} There will always be at least some contribution from non-local variations of $S_{\rm out}$ that will have to be included if the coefficient of the $\delta$ -function is small enough in magnitude. This will not turn out to be important for our analysis below, but it is something that one has to keep in mind and deserves further study.

\section{Review of the Problem}\label{sec-problem}

In this section we will review the result of Fu, Koeller and Marolf~\cite{Fu:2017lps}. Since we are treating gravity as an effective field theory, we must allow all possible terms in the action. For this analysis the important ones are the standard Einstein--Hilbert term and the Gauss--Bonnet term:
\begin{align}
I &= \frac{1}{16\pi G}\int d^dx\sqrt{g}\left[ R+\gamma \ell^2 \left((R_{\mu\nu\sigma\rho})^2 - (R_{\mu\nu})^2 + R^2\right)\right].
\end{align}
Here $\ell$ is the cutoff scale of the effective theory. The Gauss--Bonnet coupling $\gamma$ is thus defined to be dimensionless.\footnote{This is a different normalization from~\cite{Fu:2017lps}.} It will be important for us that $\gamma$ is at most order-one in this normalization, and it has been shown that this is in fact necessary to preserve causality in the effective field theory~\cite{Camanho:2014apa}.\footnote{In light of the present analysis, one could also say that this is a consequence of the QFC.} In the semiclassical regime where the Fu et al. scenario takes place, the Weyl curvature length scale $L$ satisfies $L \gg \ell$ and the equations of motion only need to be solved perturbatively in $\ell/L$. To that effect, we write the equations of motion as
\be\label{eom}
R_{\mu\nu} = \frac{\gamma \ell^2}{d-2}(C_{\eta\rho\sigma\tau})^2g_{\mu\nu} - 2\gamma \ell^2 C_{\mu\rho\sigma\tau}C_\nu^{~\rho\sigma\tau} + O\!\left(\frac{\ell^4}{L^6}\right).
\ee
Here $C_{\mu\rho\sigma\tau}$ is the Weyl tensor.

To check the QFC, we use the well-known formula for $S_{\rm grav}$ in Einstein--Gauss--Bonnet gravity~\cite{Jacobson:1993xs}:
\be
S_{\rm grav} = \frac{1}{4G}\int_\Sigma d^{d-2}y \sqrt{h}\left(1+ 2\gamma \ell^2 R_\Sigma\right).
\ee
Here $R_{\Sigma}$ is the scalar curvature of the induced metric $h_{ab}$ on $\Sigma$. From this formula one can compute the coefficient of the $\delta$-function contribution to $\delta\Theta[\Sigma,y]/\delta \Sigma(y')$, which we label $Q(y)$. The leading contribution to $Q$ comes from $\dot\theta$, which by the Raychaudhuri equation is $\dot\theta = -\theta^2/(d-2) - \sigma^2 -R_{\mu\nu}k^\mu k^\nu$. Other contributions to $Q$ will be suppressed by the cutoff scale, and the gravitational equation of motion tells us that $R_{\mu\nu}$ is suppressed as well. Then clearly $Q<0$ unless we choose to evaluate it at a point $p$ on $\Sigma$ such that $\theta|_p = \sigma|_p = 0$. In that case, one can show that~\cite{Fu:2017lps}
\be
Q|_p = 2\gamma\ell^2\left(C_{kabc}C_k^{abc} - 2C_{kba}^{~~~b}C_{kc}^{~~ac}\right) + O\!\left(\frac{\ell^4}{L^6}\right).
\ee
We are using the indicies $a,b,c$ to denote directions tangent to $\Sigma$, and the $k$ index means contraction with the null normal $k^\mu$. One may decompose the Weyl tensor into algebraically independent components,
\be
C_{kabc} = h_{ab}v_c - h_{ac}v_b + T_{abc},
\ee
where $T_{(abc)} = T_{a(bc)} = T^a_{~ba}=0$~\cite{Coley:2009is}, so that 
\be
Q|_p = 2\gamma\ell^2\left(T_{abc}T^{abc}  - 2(d-3)(d-4)v_cv^c\right) + O\!\left(\frac{\ell^4}{L^6}\right).
\ee
Clearly $Q$ is not of definite sign at the point $p$, and by considering different geometries can be made positive for any $\gamma\neq 0$. This is the conundrum posed by Fu et al., and in the next section we will see how to resolve it and rescue the QFC.

\section{Fixing the Problem}\label{sec-solution}

We will now argue that the problem of the previous section can be avoided if we smear the QFC over some $\ell$-sized region on $\Sigma$. As discussed in Sec.~\ref{sec-QFC}, the reason is that the quantity $Q$ should be thought of as a surface operator and an appropriate smearing is required to define its expectation value in effective field theory. For this reason will will adopt the notation $\langle Q\rangle$ to denote the smeared $Q$.

Let us chose our coordinates $y$ on $\Sigma$ such that the point $p$ is at $y=0$ and coordinate values of $y$ approximate proper distance along $\Sigma$, so that $\ell$ is very small in $y$-units. Then in a neighborhood of $p$ that is $\ell$-sized, we can approximate all geometric quantities by the first term in their Taylor series about $p$. Since the expansion and shear vanish at $p$, we have\footnote{Since $\theta$ and $\sigma_{ab}$ vanish at $p$, covariant derivatives can be replaced by partial derivatives, which we do for notational simplicity.}
\be
\theta(y) = y^a\partial_a\theta|_p,~~~\sigma_{ab}(y) = y^c\partial_c\sigma_{ab}|_p
\ee
Using these expressions, we can write $Q$ in this neighborhood as (dropping some irrelevant terms)
\begin{align}
Q(y) =& -\frac{(y^a\partial_a\theta|_p)^2}{d-2} - (y^c\partial_c\sigma_{ab}|_p)^2 \nonumber\\
&+2\gamma\ell^2\left(C_{kabc}C_k^{abc} - 2C_{kba}^{~~~b}C_{kc}^{~~ac}\right) + O\!\left(\frac{\ell^4}{L^6}\right).\label{Q}
\end{align}
The two terms in the first line are manifestly nonpositive, and even though they vanish at $p$ they will contribute to the smeared $\langle Q \rangle$. In the remainder of this section we will demonstrate that either those terms contribute to $\langle Q \rangle$ at order $\ell^2/L^4$ to save the QFC, or the whole expression vanishes at order $\ell^2/L^4$. In either case the potential violation will have been removed.


Naively one might think that by intelligently choosing the shape of $\Sigma$ near $p$ all of the derivatives $\partial_a\theta$ and $\partial_c\sigma_{ab}$ could be made to vanish at $p$ for any ambient geometry. However, this is not the case. The curvature of the ambient spacetime poses an obstruction, which can be easily seen in the Codazzi equation evaluated at $p$ (where $\theta$ and
$\sigma_{ab}$ are assumed to vanish)\footnote{Again, covariant derivatives can be replaced by partial derivatives because the quantities vanish at $p$.}:
\be\label{codazzi}
\partial_cK_{ab}^{(k)}|_p - \partial_bK_{ac}^{(k)}|_p = R_{kabc}|_p.
\ee
Here $K_{ab}^{(k)}$ is the extrinsic curvature of $\Sigma$ in the $k$-direction. The trace of $K_{ab}^{(k)}$ is $\theta$, and the trace-free part is $\sigma_{ab}$. Expanding $R_{kabc}$ in terms of its independent components using the notation of the previous section, we find
\begin{align}
R_{kabc} &= C_{kabc} -\frac{1}{d-2}\left(R_{kc}h_{ab} - R_{kb}h_{ac}\right)\\
&= \left(v_c - \frac{R_{kc}}{d-2}\right)h_{ab} - \left(v_b - \frac{R_{kb}}{d-2}\right)h_{ac} + T_{abc}.
\end{align}
From the definition of $K_{ab}^{(k)}$, one can see that~\eqref{codazzi} implies that some of the components of $\partial_c\sigma_{ab}$ will necessarily be at least as large as $T_{abc}$, which is of order $1/L^2$. So the potential QFC violation is ruled out unless $T_{abc} =0$ (for simplicity we are only allowing the binary choice $T_{abc} \sim 1/L^2$ or $T_{abc} = 0$, but this is not important). A similar statement applies for the derivatives $\partial_a\theta$ unless we tune the curvatures so that $v_a = R_{ka}/(d-2)$. It remains to consider that case.


Suppose then that $R_{kabc}|_p=0$, and so we can choose a surface $\Sigma$ such that even after integrating \eqref{Q} over an $\ell$-sized region it would still be apparently dominated by the term proportional to $\gamma$. We would have
\begin{align}
\langle Q \rangle &= -4(d-2)(d-3)\gamma\ell^2 v_cv^c \nonumber+ O\!\left(\frac{\ell^4}{L^6}\right)\\
&= -4\frac{d-3}{d-2}\gamma\ell^2 R_{kc}R_k^c + O\!\left(\frac{\ell^4}{L^6}\right).
\end{align}
The equations of motion, \eqref{eom}, tell us that $R_{kc} \sim \ell^2/L^4$! So our problem term is now subleading compared to terms we have dropped, and thus we cannot claim a QFC violation. Let us catalogue a few of the now-important terms that cannot be ignored.

First of all, even though we engineered the first derivatives of $\theta$ and $\sigma_{ab}$ to vanish at $p$, they could still have nonzero second derivatives. Appropriately smeared, those terms would give a negative contribution to $\langle Q \rangle$ of order $\ell^4/L^6$ unless they, too, can be suppressed. Then there are other geometric terms that we dropped from, e.g., the equations of motion, that would also contribute at order $\ell^4/L^6$. All of those would have to be analyzed and dealt with before the potential violation from the term above becomes relevant again. We leave a systematic study of those issues, and further tests of the QFC coming from geometric terms, for future work.

Beyond the geometric contributions to $d\Theta/d\lambda$ coming from $S_{\rm grav}$, there are also the contributions from $S_{\rm out}$ to consider. To begin with, $S_{\rm out}$ will contribute to the $\delta$-function coefficient of $\delta\Theta[\Sigma,y]/\delta \Sigma(y')$ in a way that scales like $\ell^{d-2}/L^{d-4}$. In five dimensions this is already enough to overpower our suppressed geometric terms, and may be enough in higher dimensions depending on the extent of the suppression. The contribution of $S_{\rm out}$ to the $\delta$-function term of the QFC is very difficult to compute in general, which is the primary reason why the QFC is a conjecture and not a theorem.\footnote{The free-field QNEC proof in~\cite{Bousso:2015wca} was essentially a calculation of these $\delta$-function contributions from $S_{\rm out}$.} It would be very interesting to see if there are situations where it could be made positive, resulting in a possible tension with the QFC.

Finally there are the nonlocal terms in $d\Theta/d\lambda$, which one has to include since, as discussed in the paragraphs following \eqref{QFC}, the profile $f(y)$ cannot be made to have support smaller than the cutoff scale. Like the $\delta$-function contributions of $S_{\rm out}$, the nonlocal contributions are difficult to compute. However, at least here we have the advantage of being able to use strong subadditivity to say that they are negative. It is unclear how the nonlocal terms coming from $S_{\rm out}$ compare in magnitude with the $\delta$-function contributions from $S_{\rm out}$ in the limit where the support of $f$ becomes of order the cutoff size. It is reasonable to suppose that they are negligible, but this is another area for future work.

\section{Discussion}\label{sec-disc}

We have seen that the formulation of the QFC requires a careful treatment of the geometric terms as surface operators on $\Sigma$, meaning that they have to be smeared over some region in order to be well-defined. This is related to the observation of Flanagan and Wald~\cite{Flanagan:1996gw} that the Averaged Null Energy Condition (ANEC), which is one of the consequences of the QFC, can be violated unless one employs a cutoff-scale smearing.\footnote{I thank Netta Engelhardt for drawing my attention to that result.} It is clear, then, that one cannot try to formulate semiclassical quantum gravity statements without giving a smearing prescription to make the operators well-defined.

To conclude, we comment on how this same issue of operator smearing is important for Entanglement Wedge Nesting (EWN) in AdS/CFT. EWN is the statement that quantum extremal surface~\cite{Ryu:2006bv, Hubeny:2007xt,Engelhardt:2014gca} associated to a region on the boundary must move in a spacelike way when the boundary region grows or shrinks. EWN can be viewed as a consequence of the QFC in the bulk~\cite{Wall:2012uf, Engelhardt:2014gca, Akers:2016ugt} or as a requirement for the consistency of subregion duality, for which there is ample evidence~\cite{Czech:2012bh, Jafferis:2015del, Dong:2016eik}. EWN and the QFC are also related in that they can both be used to prove the QNEC~\cite{Bousso:2015mna, Koeller:2015qmn, Akers:2016ugt}. There is likely an even deeper connection between the two, as evidenced by the the following (to be analyzed in more detail in forthcoming work~\cite{toappear}).

If one attempts to check that EWN is satisfied when the boundary is curved and has $d\geq  5$ spacetime dimensions, one finds that it can naively be violated when the bulk theory contains a Gauss--Bonnet coupling. This setup is precisely analogous to the one found by Fu et al. for the pointwise QFC violation. In fact, the formulas themselves are remarkably similar. The resolution to the EWN violation is also analogous: the ``spacelike" condition we are trying to verify in the bulk is just the sign of a certain bulk surface operator expectation value, and that operator must be smeared over the bulk cutoff scale in order to be well defined. This prescription saves EWN just like the smearing prescription given above saves the QFC. The extreme similarity leads one to speculate that EWN and the QFC can perhaps be identified in a braneworld scenario or similar setup.


\begin{acknowledgments}
I would like to thank Z.~Fu, J.~Koeller, and D.~Marolf for sharing a draft of \cite{Fu:2017lps} prior to publication. In addition, thank you to C.~Akers, V.~Chandrasekaran, R.~Bousso, N.~Engelhardt, A.~Levine, A.~Shahbazi Moghaddam and A.~Wall for stimulating discussions on its contents. My work is supported in part by the Berkeley Center for Theoretical Physics, by the National Science Foundation (award numbers 1521446, and 1316783), by FQXi, and by the US Department of Energy under contract DE-AC02-05CH11231.
\end{acknowledgments}


\bibliographystyle{utcaps}
\bibliography{QFCSmearingLetter}

\end{document}